\documentclass[aps,amsfonts,amsmath,prd,preprint,nofootinbib]{revtex4}
\usepackage{epsf}

\newcommand{\beq}{\begin{equation}}
\newcommand{\eeq}{\end{equation}}

\begin{document}

\preprint{MIT-CTP-3800}

\title{Eternal inflation, bubble collisions, and the persistence
of memory}

\author{Jaume Garriga}
\affiliation{Departament de Fisica Fonamental, Universitat de Barcelona, 
Marti i Franques 1, 08028 Barcelona, Spain}
\author{Alan H. Guth}
 \affiliation{Center for Theoretical Physics,
     Laboratory for Nuclear Science and Department of Physics,
     Massachusetts Institute of Technology, Cambridge,
     MA 02139, USA.}
\author{Alexander Vilenkin}
 \affiliation{Institute of Cosmology, Department of Physics and Astronomy\\
     Tufts University, Medford, MA 02155, USA.}

\begin{abstract}

A ``bubble universe'' nucleating in an eternally inflating false
vacuum will experience, in the course of its expansion,
collisions with an infinite number of other bubbles.  In an
idealized model, we calculate the rate of collisions around an
observer inside a given reference bubble.  We show that the
collision rate violates both the homogeneity and the isotropy of
the bubble universe.  Each bubble has a center which can be
related to ``the beginning of inflation'' in the parent false
vacuum, and any observer not at the center will see an
anisotropic bubble collision rate that peaks in the outward
direction.  Surprisingly, this memory of the onset of inflation
persists no matter how much time elapses before the nucleation of
the reference bubble.

\end{abstract}

\maketitle

\section{introduction}

In a theory with different metastable vacua, such as the
landscape of string theory \cite{BP,Susskind}, the process of
eternal inflation may lead to a ``multiverse'' where different
regions of space-time are occupied by different vacua. A region
occupied by vacuum A may spawn regions of other vacua B,C,D...
which are adjacent in field space.  Daughter regions will in turn
spawn their offspring, and so on, producing an infinite tree of
``pocket universes.''

Vacua can evolve by a mix of semiclassical tunneling and
stochastic evolution, but in this paper we will consider only
tunneling.  In that case, bubbles form by random nucleation and
then start to expand into the parent vacuum with constant
acceleration.  The interior of the growing bubble has the
geometry of an open FRW universe
\cite{CdL}, and (assuming that a short period of slow roll
inflation flattens it out to satisfaction) we may entertain the
possibility that we live in one of such ``bubble universes''
\cite{open}.

The above description, however, is incomplete because it ignores
collisions with other bubbles.  Collisions may be quite rare if
the nucleation rates are small. Nonetheless, a bubble expands for
an infinite amount of time, and will collide and merge with an
infinite number of other bubbles, forming an ever growing
``cluster''.

We should therefore reevaluate the naive picture of a smooth FRW
universe on large scales. In particular, we may ask what fraction
of the FRW universe remains unaffected by collisions. As we shall
see, only a set of measure zero remains unaffected. If this is
supposed to describe our local universe, there seems to be reason
for concern. Why haven't we seen any collisions yet? Do we have
much time left until we are blown away by a collision? What is
the expected distance to the nearest point in our FRW time slice
which {\em has} already been hit by another bubble? The purpose
of the present paper is to explore some of these issues.

In the course of this investigation we have stumbled upon a
rather remarkable result. We find that the rate of collisions
around a typical observer in the bubble universe is {\em
anisotropic}, and the origin of this anisotropy is related to the
beginning of false vacuum inflation.

A metastable inflationary de Sitter phase can be eternal to the
future, but not to the past.  More precisely, the inflating
region of spacetime is geodesically past-incomplete \cite{BGV},
and therefore some initial conditions must be specified on the
past boundary of this region. For instance, we may posit that at
some initial time, a given large region of space is in false
vacuum. The congruence of geodesics normal to the initial time
surface defines a preferred frame, with respect to which
velocities can be defined.

The standard lore is that this preferred frame is not important,
and that the initial conditions are soon forgotten. This may be
true for geodesic observers in the false vacuum phase, whose
velocity with respect to the preferred congruence redshifts
exponentially with time. For those ``co-moving'' observers,
memory of the initial surface is lost as we push the surface far
away into the remote past. However, when a bubble of a new phase
forms in the original false vacuum, the congruence of observers
in the FRW open universe includes observers with velocities
arbitrarily close to the speed of light relative to the preferred
congruence. Such individuals are not ``far'' from some points on
the initial surface, and may have a chance to detect {\em some}
information about the beginning of inflation.

The plan of the paper is the following. In Section II we describe
the geometry and statistics of collisions onto a given reference
bubble. In Section III, we consider the anisotropies in the
bubble distribution as seen by observers inside a reference
bubble. The origin of these anisotropies is further discussed in
Section IV in a simplified context, where the observer is in
false vacuum rather than inside the bubble.  Section V is devoted
to conclusions. Some technical details are left for the
appendices.

\section{Collisions onto a reference bubble}

Here, we shall derive some basic results concerning the
distribution of collisions impinging on a given reference bubble.
We begin with a description of the basic setup we shall consider.

It is convenient to use flat de Sitter coordinates to describe
the background inflating false vacuum,
\beq
ds^2 = dt^2 - e^{2Ht}(dr^2+r^2d\Omega^2).
\eeq
To simplify the equations, we shall choose units so that
\beq
H=1.
\eeq
To simplify matters further, we shall assume that the vacuum
energy density inside the bubbles is nearly the same as that
outside (at least for a sufficiently long time after bubble
nucleation) and that the gravitational effect of bubble walls is
negligible. Then the metric is
\beq
ds^2 = dt^2 - e^{2t}(dr^2+r^2d\Omega^2)
\eeq
in the entire spacetime region of interest.

It is also useful to consider the ``embedding'' of de Sitter
space as a timelike hyperboloid of unit radius, in a 5
dimensional Minkowski space, whose rectangular coordinates are
labeled $V,W$ and $\vec X =(X,Y,Z)$:
\begin{equation}
\vec X^2 + W^2 - V^2 = 1.\label{theh}
\end{equation} 
These are related to the flat chart coordinates $t$ and $\vec
x=(x,y,z)$, ($r\equiv |\vec x|$), through
\begin{equation}
W-V = e^{-t}-e^{t} r^2,\quad W+V = e^t,\quad \vec X= e^t \vec x.
\label{emb}
\end{equation}
Let us consider a ``reference'' bubble that nucleates at $t=r=0$
(we shall sometimes call this ``our'' bubble).  In the embedding
coordinates, this corresponds to
\begin{equation}
W=1, \quad V=\vec X = 0.
\end{equation} 
The bubble geometry is symmetric under boosts which have the
nucleation event as a fixed point.  These form an $O(3,1)$ group
of isometries.

The interior of the bubble (or more precisely, the interior of
the light cone from the nucleation event) is described by the
line element
\beq
ds^2=d\tau^2-\sinh^2\tau(d\xi^2+\sinh^2\xi d\Omega^2).
\eeq
The coordinates $(t,r)$ and $(\tau,\xi)$ are related by
\begin{eqnarray}
e^t &=& \cosh \tau + \cosh\xi \sinh\tau, \label{deco}\\ e^t r &=& \sinh \xi \sinh
\tau.\label{fto}
\end{eqnarray}
We set the initial condition that there are no bubbles at some
$t=t_i$. This breaks the residual $O(3,1)$ invariance of the
bubble, which is responsible for the homogeneity and isotropy of
our FRW universe. Consequently, not all observers who live in the
open FRW universe will see the same.

\subsection{Collisions around the observer at $\xi=0$.}

Let us now concentrate on the distribution of collisions around
the point $\xi=\xi_{\rm obs}=0$, which is at the origin of the
open FRW hyperboloid.  We leave the consideration of typical
observers, who live far away from the origin, to the next
Subsection and to Section III.

The distribution around $\xi_{\rm obs}=0$ will of course be
isotropic.  In particular, we shall be interested in the typical
distance at which we might expect the nearest collision.  In what
follows, we shall calculate the probability $P(\xi,\tau)$ that
{\em no} collisions with other bubbles have affected a spherical
region of radius $\xi$ around the origin, on a hypersurface of
constant $\tau$.

\begin{figure}[ht]
\begin{center}\leavevmode  %
\epsfxsize=16.5cm\epsfbox{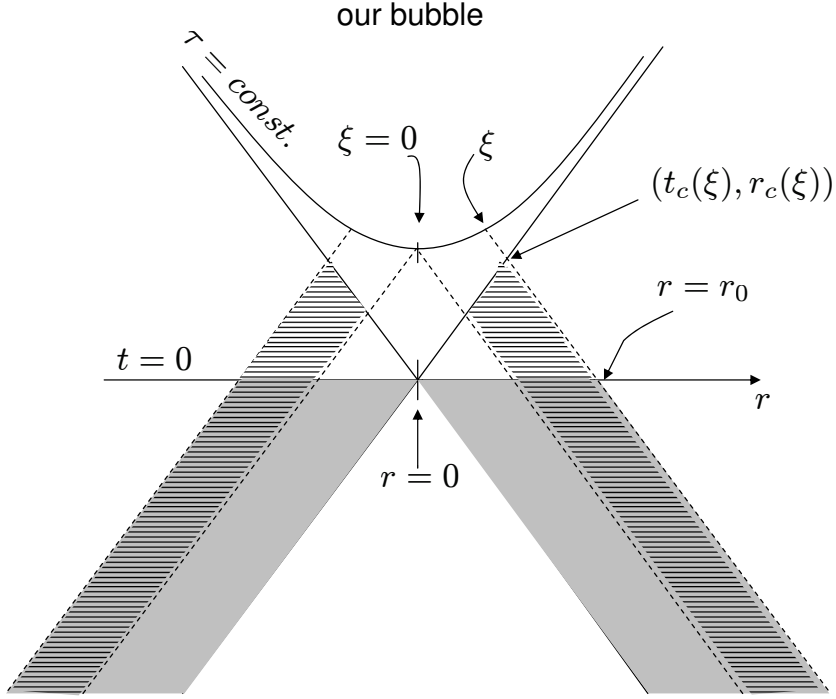}
\end{center}
\caption{Region in grey shade, ${\cal V}_4(r_0)$ can nucleate bubbles 
which will be within the distance $r_0$ from the origin (on the
flat surface $t=0$). Here, we assume that the point $r=t=0$ is
still in false vacuum.  Region in horizontal stripes,
$V_4(\xi,\tau)$, can nucleate bubbles which will collide with
ours, and which will be visible within a coordinate radius $\xi$
from the origin (on the $\tau=const.$ hypersurface). Here, we
assume that the center point $\xi=0$ has not yet been hit by any
bubble.  }
\label{1j}
\end{figure}

As a warm up exercise, let us consider the following question.
Assuming that the point $t=r=0$ is still in false vacuum, what is
the probability that some bubble will have hit the surface $t=0$
at some $r\leq r_0$?. The relevant quantity is the 4-volume
inside the past light-cone of the the circle $r < r_0$ minus the
4-volume inside the past light cone of the origin $r = 0$ (see
Fig. 1).  The physical radius of the past light cone from the
origin is $R_0(t)=1-e^t$ ($t<0$), whereas the radius of the past
light cone from the circle of radius $r_0$ is $R_{1}(t) = R_0(t)
+ r_0\ e^t$. The volume of the grey shaded region in Fig.
\ref{1j} is thus given by
\begin{equation}
{\cal V}_4(r_0, t_i)= {4\pi\over 3} \int_{t_i}^0 (R_{1}^3-R_0^3)\ dt = 
{2\pi\over 9} (6 r_0 + 3 r_0^2 + 2 r_0^3)+O(e^{t_i}r_0). \label{v4dist}
\end{equation} 
Note that this is finite (for finite $r_0$) even in the limit
when the initial surface is pushed all the way to $t_i\to
-\infty$. For a region of a Hubble size, we have
$$ 
{\cal V}_4(1, t_i\to -\infty) =
{22 \pi \over 9}, 
$$ 
Hence, the probability of having a bubble one Hubble distance
away from ours at $t=0$ is of order of the nucleation rate per
unit volume $\lambda$, which we shall assume to be small
($\lambda\ll 1$ in the units where $H=1$).  The distance to the
nearest bubble at $t=0$ can be estimated from the condition
$\lambda {\cal V}_4 \sim 1$. From Eq.~(\ref{v4dist}), this
distance will be of the order $r_0 \sim
\lambda^{-1/3}$.

Next, we may consider the same question on a $\tau=const.$
hyperboloid. Past directed radial null rays define a mapping
between points on the hyperboloid at some distance $\xi$ and
points on the ``bubble cone,'' by which we mean the future
light-cone from the nucleation event. The bubble cone is given by
the equation
\begin{equation}
r(t) = 1-e^{-t}. \quad (t>0). \label{flc}
\end{equation}
The flat chart coordinates of points on the hyperboloid can be
found from Eq.~(\ref{fto}).  The past directed outward radial
null geodesic from a point $(\xi,\tau)$ is given by
\begin{equation}
r(t)= e^{-t} + F(\tau,\xi), \label{rf}
\end{equation}
where
\begin{equation}
F={\sinh \xi\sinh \tau -1 \over \cosh \tau +\cosh\xi
\sinh\tau}.\label{Fen}
\end{equation}
Note in particular that for $\xi=0$, we have $F=-e^{-\tau}$,
whereas for large $\xi$ and $\tau$, we have
\begin{equation}
F(\xi,\tau)\approx 1-2 e^{-\xi}+2 e^{-2 \xi}-4 e^{-\tau-\xi} .\quad
(\xi,\tau \gg 1)\label{assymtF}
\end{equation}
The intersection of the null geodesic (\ref{rf}) with the bubble
cone (\ref{flc}) is at
\begin{equation}
e^{-t_c} = {1 - F\over 2}, \quad  r_c= {1 + F \over 2}. \label{tcrc}
\end{equation}
(Incidentally, the intersection with the plane at $t=0$ is at
$r_0 = 1 + F < 2.$)

It will be very convenient to use Eq.~(\ref{rf}) as the
definition for a change of variables, to replace $r$ in favor of
$F$. The different values of $F$ can be thought of as labeling
the different past directed outward radial null rays emanating
from a given $\xi$, on a $\tau=const.$ hyperboloid, as indicated
by Eq.~(\ref{Fen}). More explicitly, we introduce
\begin{equation}
a\equiv e^t,\quad F\equiv r- e^{-t},
\end{equation}
in terms of which the metric reads
\begin{equation}
ds^2= 2da dF - a^2 dF^2 - (1+aF)^2 d\Omega^2.\label{af}
\end{equation}
This form of the metric is completely regular at $a=0$ (which
corresponds to the boundary of the flat coordinatization). In
fact, the new chart (\ref{af}) covers the {\em whole} of de
Sitter space, when we let the coordinates vary in the range
$-\infty< a <\infty$, $-\infty<F<\infty$, with the restriction
$aF>-1$. Negative values of $a$ correspond to the part of de
Sitter space not covered in a flat chart.

Let us assume that, on a given $\tau=const.$ hyperboloid, the
origin $\xi=0$ has not yet been hit by a bubble, and let us ask
what is the typical distance at which we may expect the nearest
collision.  The probability that no collisions have affected a
region of coordinate radius $\xi$ around the origin is given by
\begin{equation}
P(\xi,\tau,t_i) = e^{-\lambda V_4},\label{persistence}
\end{equation}
where $\lambda$ is the nucleation rate, and $V_4$ is the relevant
4 volume to the past of a spherical region of radius $\xi$ on the
hyperboloid. Here, we must subtract the contribution from the
interior of the past light cone of $\xi=0$ (this is because we
assume $\xi=0$ has not been hit by any bubbles), and also the
intersection with the interior of our bubble, since we assume
that no new bubbles can nucleate inside of our bubble.  We may
express this 4 volume as
\begin{equation}
V_4(\xi,\tau,t_i) = 4\pi \int e^{3t} r^2 dr dt = 4 \pi \int e^{3 t}
(e^{-t} + F)^2 dF dt = 4\pi \int (1+a F)^2 da dF.
\label{v44}
\end{equation}

The evaluation of $V_4$ is simplest in the $(F,a)$ coordinates.
This is because the light cones are bound by null geodesics with
$F=const.$ Hence, the integration limits for $F$ in (\ref{v44})
are independent of $a$,
$$ 
-e^{-\tau} < F < F(\tau,\xi).  
$$ 
The lower limit corresponds to the center of the hyperboloid
$\xi=0$. The integration range for $t$ depends on $F$: $\ t_i < t
< t_c(F),$ where $t_c$ is given in (\ref{tcrc}). Here we are
approximating the boundary of our bubble by the ``bubble cone,''
which is justified when the size of bubbles at the time of
nucleation is much less than the Hubble radius. In terms of
$a=e^t$, the range is given by
$$
a_i < a < 2/(1-F).
$$
Performing the double integral, we have
\begin{equation}
V_4(\xi,\tau,t_i) = {8\pi\over 3} \left[ {2F\over
(1-F)^2}-\ln(1-F)+O(a_iF)\right]_{F=-e^{-\tau}}^{F(\xi,\tau)},
\label{v444}
\end{equation}
where the upper limit is given by (\ref{Fen}). Note that the
4-volume takes a finite value in the limit when the initial
surface is pushed into the remote past $a_i\to 0$.

Defining $V_4(\xi,\tau)\equiv\lim_{t_i\to
-\infty}V_4(\xi,\tau,t_i)$, we find
\begin{equation}
V_4(\xi,\tau) = {8 \pi \over 3} \left\{e^\xi \sinh \xi \,
     \tanh^2 \left({\tau \over 2}\right) + \ln \left[ 1 + {1\over 2} (1-e^{-\tau})
     (e^\xi - 1) \right] \right\}.\label{alan}
\end{equation}
In particular, for $\tau\gg 1$, we have
\begin{eqnarray}
V_4(\xi,\tau) &\approx& 4\pi\xi + O(\xi^2) \quad(\xi\ll 1),\label{v4s}\\ 
V_4(\xi,\tau) &\approx& {4\pi\over 3} e^{2\xi} + O(\xi)\quad(\xi\gg 1).
\label{v4xi}
\end{eqnarray}
The typical coordinate distance to the nearest bubble is found
from $\lambda V_4 \sim 1$, and for small nucleation rates,
$\lambda\ll 1$, it is of order
\begin{equation}
\Delta\xi \sim \ln(1/ \lambda).\label{estimate}
\end{equation} 
Note that this is independent of $\tau$. This is to be expected,
since (for $\tau\gtrsim 1$) the products of collision are
``frozen in'' with the expansion of the universe during the
period of inflation inside the bubble. Here, for simplicity, we
have approximated this period as a de Sitter phase with the same
expansion rate as the false vacuum. In a more realistic case we
should consider a period of slow roll inflation, followed by a
standard decelerating phase. If this phase leads to a Minkowski
vacuum, all products of collision would gradually fall within the
horizon, and eventually hit {\em any} observer inside the bubble.
(This is explicitly shown at the end of Appendix 1, where we
calculate the rate at which an observer in Minkowski vacuum would
see new bubbles falling into his field of view.) On the other
hand, our own universe does not seem to be approaching a
Minkowski vacuum. Rather, it appears to be accelerating again,
and collisions with other bubbles may forever remain hidden
behind the cosmological horizon.

The curvature radius of the FRW universe corresponds to
$\Delta\xi\sim 1$. Current observational bounds on the spatial
curvature imply \cite{Spergel} $(a_0 H_0)^{-2}=|1-\Omega|\lesssim
10^{-2}$, where the subindex $0$ indicates the present time.  The
coordinate size of the observable universe is given by
$\Delta\xi_0 \sim (a_0 H_0)^{-1}$. Hence, the observer at the
center of the hyperboloid $\xi_{\rm obs}=0$ can only see out to
coordinate distances $\Delta\xi_0\lesssim 10^{-1}$. Using
(\ref{persistence}) and (\ref{v4s}), the probability that this
observer may see any collisions at all is given approximately by
$$ 1-P(\Delta\xi_0) \approx 1-e^{-4\pi\lambda
\Delta\xi_0}\lesssim \lambda\ll 1. 
$$ 
This issue will be discussed further in Section III, and also in
Appendix 1.  In the Appendix we show that the asymptotic
expressions (\ref{v4s}) and (\ref{v4xi}) are valid also for the
relevant four volume in the neighborhood of any observer (at
$\xi_{\rm obs}\gg 1$) who has not yet been hit by a bubble.  This
means that we are very unlikely to be hit by a bubble at any time
in the future.

The expected physical distance to the collision nearest to
$\xi=0$, at the time when inflation ends, can be estimated as
\begin{equation}
d= Z \ln(1/\lambda),
\label{dZ}
\end{equation}
where $Z\sim \sinh\tau_f$ is the slow-roll expansion factor
($\tau_f$ is the time at the end of inflation inside the bubble). 
Here we have worked in a simplified scenario where H is the same
in false vacuum as it is in the period of slow-roll inflation
inside the bubble. In the general case, the expression for $d$
will be more complicated, but the dependence on $Z$ and $\lambda$
will be similar.

\subsection{Fractal dimension of the bubble universe.}

The world-line of a given point $(\theta,\phi)$ on the wall of
the reference bubble will sooner or later be hit by other
bubbles, with probability equal to one.  This follows from the
fact that there is a finite probability per unit time $t$ for
this worldline to be hit by other bubbles (see Section
\ref{moving} for a rigorous derivation of this statement). 
Hence, the fraction of the bubble wall {\bf area} that makes it
to future infinity without collisions is a set of measure zero.

The effect of collisions onto the reference bubble propagates
into the open FRW universe.  On the hyperbolic slice
$\tau=const.$, each collision affects a ``wedge'' shaped region,
whose tip points towards the origin and which extends all the way
to $\xi\to \infty$, spanning a finite asymptotic solid angle at
spatial infinity. Let us denote by $\xi_c$ the distance from the
tip of the wedge to the origin, and let us calculate the
asymptotic angular size of the wedge, $\theta_w$, as a function
of $\xi_c$.

It is convenient, again, to use the coordinates $a,F$ of the
chart (\ref{af}). A bubble that nucleates at time $a$ has
asymptotic co-moving radius $1/a$ at future infinity. A bubble
nucleating at point $F$ will have its center displaced a distance
$r=F+a^{-1}$ from the origin, and its boundary at future infinity
will have co-moving cartesian coordinates satisfying:
\begin{equation}
x^2+y^2+(z-F-a^{-1})^2 = a^{-2}.
\end{equation}
(Here we have assumed that the center is on the $z$ axis).  The
future boundary of the reference bubble (which nucleates at
$t=r=0$) is given by
\begin{equation}
x^2+y^2+z^2=1.
\end{equation}
The intersection of both is at
\begin{equation}
z=\cos\theta_w ={1\over 2}{a+2F+aF^2 \over 1+aF}\label{ang}
\end{equation}
Note that $\theta_w$ corresponds to the asymptotic angular size
of wedges corresponding to bubbles nucleated at $(a,F)$. 

On a hyperboloid with $\tau\gtrsim 3$, large $\xi$ corresponds to
$1-F\approx 2e^{-\xi}(1+\cosh\tau)/\sinh\tau\approx 2 e^{-\xi}$. 
From (\ref{v44}), the bubbles which affect this region nucleate
in a four-volume dominated by the largest possible
$a$, $$ a\sim {2\over 1-F}\sim e^{\xi}.  $$ From (\ref{ang}), The
typical angular size of wedges whose tip is at $\xi_c\gg 1$ is
therefore
\begin{equation}
\theta_w(\xi_c)\sim e^{-\xi_c}\ll 1.\label{tw}
\end{equation}

Let us denote by $\Omega_u(\xi)$ the solid angle which remains
unaffected by collisions out to a distance $\xi$ from the origin.
As $\xi$ is increased, more and more wedges accumulate, each one
removing a solid angle $2\pi [1-\cos\theta_w]$ from $\Omega_u$.
The loss of solid angle due to wedges whose tip is in the
interval $d\xi$ is given by
\begin{equation}
d\Omega_u(\xi) \approx -2\pi [1-\cos\theta_w]\lambda dV_4 \sim
-\lambda e^{-2\xi}dV_4,
\label{do}
\end{equation}
where $dV_4$ is the region of four volume to the past of the
interval $d\xi$, in the unaffected portion of the sphere. Here,
we are making the approximation that wedges do not overlap with
each other. Also, we shall assume that a wedge whose tip is at
$\xi_c$ depletes a solid angle equal to the corresponding
asymptotic value (at $\xi\to \infty$) for all $\xi > \xi_c$.  The
four volume is approximately given by Eq.~(\ref{v44}), where
$4\pi$ is replaced by $\Omega_u(\xi)$. Using (\ref{ang}), we have
\begin{eqnarray}
2\pi [1-\cos\theta_w]dV_4 &=&\pi \Omega_u(\xi)
\left[\int_{a=0}^{2/(1-F)} (1-F)[2-a(1-F)](1+aF)da\right]
dF\nonumber\\ &=&{2\pi (3-F)\over3(1-F)} \Omega_u(\xi)\ dF\approx
{4\pi\over 3} \Omega_u(\xi) d\xi,\label{long}
\end{eqnarray} 
where in the last step we have used $F\approx 1-2 e^{-\xi}$. 

Combining Eq.~(\ref{long}) with (\ref{do}), the solid angle which
remains unaffected by collisions as a function of $\xi$ is given
by
\begin{equation}
\Omega_u(\xi) \approx 4 \pi e^{-\kappa \xi}, \quad (\xi\gg 1),
\label{omegau}
\end{equation}
where $\kappa\approx (4\pi/3)\lambda$. The volume element on a
spatial section of an open FRW is given by $dV^{FRW}= 4\pi
\sinh^2\xi d\xi$. The volume which is {\em unaffected} by bubbles
\begin{equation}
d V = \Omega_u(\xi) \sinh^2\xi d\xi \sim 2\pi e^{(2-\kappa)\xi} d\xi,
\label{dv}
\end{equation}
is therefore unbounded and dominated by large distances from the
origin (assuming, of course, that $\kappa\ll 1$).  Nevertheless,
the unaffected {\em volume fraction} tends to zero in the limit
$\xi \to \infty$.  The unaffected volume fraction will be
recalculated in the next section, using a method which is exact
in the context of our idealized model.

The fact that the unaffected part of the volume is a set of
measure zero might seem to be reason for concern. Nevertheless,
if we pick a random point in the unaffected region, then the
closest hit is likely to be quite far away. Indeed, the number of
bubbles hitting in the interval $d\xi$ is given by $dN= \lambda
dV_4$. Using (\ref{tw}), (\ref{long}) and (\ref{dv}) we find that
for large $\xi$,
\begin{equation}
dN \sim \lambda dV.\label{numberdensity}
\end{equation}
Let us now pick a random point in the unaffected region, at
$\xi\gg 1$, and let us choose it as the origin of open FRW
coordinates. A spherical region of radius $\Delta\xi\gg 1$ around
that point has coordinate volume $V\sim e^{2\Delta\xi}$. Using
(\ref{numberdensity}), the expected distance to the nearest
bubble will be of order
\begin{equation}
\Delta\xi \sim \ln(1/\lambda).\label{cds}
\end{equation}
This is the same result we found in the previous Subsection, for
the distribution around the privileged point at the center of
the hyperboloid. The coordinate distance (\ref{cds}) is large
enough that we should not be too concerned about the hazard of
future collisions.

Finally, let us characterize the fractal dimension of the
unaffected part of the bubble universe.  Looking outward from the
center of the hyperboloid, we see finer and finer ``wedges''
carving away the solid angle as we increase $\xi$. If we define a
smearing angle $\epsilon\equiv e^{-\xi}$, and ignore structures
smaller than $\epsilon$, the unaffected solid angle is
$\Omega_u(\xi) \approx 4 \pi
\epsilon^{\kappa}$. The number of sets of angular radius $\epsilon$
which is needed to cover this region is $n \sim
\epsilon^{{\kappa}-2}$. Thus we can think 
of
\beq
D \equiv 2 - \kappa = 2 - {4 \pi \over 3} \lambda
\eeq
as the fractal dimension of the unaffected portion of the
``celestial'' sphere at $\xi\to\infty$.  Clearly, this is also
the fractal dimension of the remaining surface of our bubble
which has not been hit by other bubbles at future infinity.

\section{Anisotropies in the distribution of collisions}

To study collisions around points which are away from the origin,
it is useful to perform a de Sitter transformation (Lorentz
transformation in the embedding space) to a new frame S$'$ where
the point of interest {\em is} at the origin. This greatly
simplifies the geometry of the relevant past light cones.

If the original point at large $\xi=\xi_{\rm obs}$ was along the
$Z$ direction, then we use the boost
\begin{equation}
V'=\gamma(V-\beta Z),\quad Z'=\gamma(Z-\beta V),\quad X'=X,\quad
Y'=Y,\quad W'=W,
\label{coordtran}
\end{equation}
with $\beta=\tanh \xi_{\rm obs}$ and $\gamma=\cosh\xi_{\rm obs}$,
in order to bring the point of interest down to $\xi_{\rm
obs}'=0$. In doing so, the initial surface $t=t_i$ gets distorted
and this is of course something we have to consider.

In the embedding coordinates, $t=t_i$ corresponds to the null
plane
\begin{equation}
W+V=e^{t_i}\equiv a_i,\label{isu} 
\end{equation}
which in boosted coordinates reads
\begin{equation}
V'+\beta Z' = \gamma^{-1} (a_i-W').
\label{isubo}
\end{equation}
We can now express this in terms of flat chart coordinates
$(t',\vec x')$ by using the standard relations (\ref{emb}).  This
leads to
\begin{equation}
\sinh t' +{1\over 2} e^{t'} r'^2 +\beta e^{t'} r' \cos\theta' = 
\gamma^{-1}(a_i-\cosh t' +{1\over 2} e^{t'} r'^2). \label{teq0}
\end{equation}
In terms of coordinates $(a',F')$ analogous to the $(a,F)$ pair
which we introduced in the previous Section [see Eq.~(\ref{af})],
the equation for the initial surface $a=a_i$ takes the more
tractable form $a'=a'_i(F',\theta';a_i)$, where
\begin{equation}
a'_i= 2 \ { a_i-\beta\gamma\cos\theta'-(\gamma-1)F'\over 1+\gamma+2
\beta\gamma F'\cos\theta' + (\gamma-1)F'^2}.\label{insur}
\end{equation} 
It should be noted that $a_i'$ can be negative for some values of
$F'$. This is not a problem. It just means that in the boosted
frame, the initial surface invades the portion of de Sitter space
not covered in the original chart. 

It is interesting to calculate the probability that the FRW
observer at $\xi'=0$ (that is, $\xi=\xi_{\rm obs}$) will be
affected by a collision with another bubble before some specified
time $\tau$. This is related to the spacetime volume available
for the nucleation of bubbles which lies in the past light cone
of the observation point $(\xi'=0,\tau)$, but which does not lie
inside our reference bubble, or inside the past light cone of the
nucleation event.  This four volume increases with proper time,
in the following way
\begin{equation}
dV_4 = \left[\int_{a'_i}^{2/(1-F')}(1+a'F')^2 da'\right] {dF'\over
d\tau} d\tau d\Omega',\label{stpt}
\end{equation}
where $d\Omega'=2\pi\ d(\cos\theta')$. The probability per unit
proper time and solid angle that our observer is hit by a bubble,
assuming that he was not already hit by
     a bubble, is given by
\beq
{dP\over d\tau\ d\Omega'}= \lambda{dV_4 \over d\tau d\Omega'}.
\eeq
From Eq.~(\ref{Fen}), $\xi'=0$ corresponds to $F'=-e^{-\tau}$,
and performing the integration over $a'$ we find
\begin{equation}
 {d V_4 \over d \tau d \Omega'} = {1 \over 3} \left[ \left(
     {\cosh \xi_{\rm obs}\, \sinh \tau + \cosh \tau - a_i \over
     \cosh \xi_{\rm obs} \, \cosh \tau +\sinh \tau - \sinh
     \xi_{\rm obs} \, \cos \theta' } \right)^3- \tanh^3 \left(
     {\tau \over 2} \right) \right].
\label{lagr}
\end{equation}
Note that the result depends both on the point of observation
$\xi_{\rm obs}$ and on the direction of observation $\theta'$.
Homogeneity and isotropy are lost, even in the limit when we push
the initial surface all the way to $t_i\to -\infty$, i.e.
$a_i=0$.

For any fixed nonzero value of $\tau$, the limit of (\ref{lagr})
as $\xi_{\rm obs}\to \infty$ is given by
\begin{equation}
{dV_4\over d\tau d\Omega'} = {1\over 3}\left[\left({\sinh\tau\over
\cosh\tau - \cos\theta'} \right)^3-
\tanh^3 \left( {\tau\over 2} \right) \right].\label{farout}
\end{equation}
The existence of this limit is good news. The number of observers
grows without bound with the distance to the center of the
hyperboloid, and thus we expect that the typical observer lives
at very large $\xi_{\rm obs}$. He or she should therefore measure
the distribution (\ref{farout}).  Also, the limit is independent
of $t_i$. This is also interesting, since it means that some
specific details about the surface of initial conditions do not
seem to matter. Nevertheless, the distribution is anisotropic,
with the minimum number of hits per unit time in the direction of
the center of our reference bubble. In this sense, memory of the
initial surface persists. For large $\tau$, the distribution
takes the simple dipole form
\begin{equation}
{dV_4\over d\tau d\Omega'} = 2(1+\cos\theta') e^{-\tau} +
O(e^{-2\tau}).
\end{equation}

Dipole anisotropies and memory of initial conditions are not
commodities one usually expects from inflation.  Although we find
this result to be rather shocking, the effect is real.  Its
origin is best understood by eliminating the complications due to
the bubble geometry, as will be discussed in the next section.
First, however, we would like to study a bit further the
implications of (\ref{lagr}).

To find the total rate $\lambda d V_4 / d \tau$ at which bubbles
will be encountered by an observer who has not previously been
hit by a bubble, one can integrate (\ref{lagr}) over solid angle. 
The result is given by
\beq
{dV_4 \over d \tau} ={4 \pi \over 3}\left[{(\cosh \xi_{\rm
     obs}\, \sinh \tau +\cosh \tau -a_i)^3 (\cosh \xi_{\rm obs}
     \, \cosh \tau +\sinh \tau ) \over (\cosh \xi_{\rm obs} \,
     \sinh \tau +\cosh \tau )^4} - \tanh^3 \left( {\tau \over 2}
     \right) \right].
\label{dVdtau}
\eeq
The dependence of this result on $\xi_{\rm obs}$ shows us the
inhomogeneity of the bubble collision rate.  For the special case
of $\tau=0$ the expression simplifies to
\beq
{dV_4 \over d \tau} ={4 \pi \over 3} (1-a_i)^3 \cosh \xi_{\rm
     obs} ,
\eeq
which is highly inhomogeneous, while for large $\tau$ the
expression becomes
\beq
{dV_4 \over d \tau} = 8 \pi \left( 1- {a_i \over \cosh \xi_{\rm
     obs}+1} \right) e^{- \tau} + O(e^{-2 \tau}) ,
\eeq
which shows that the inhomogeneity disappears when $t_i \to
-\infty$ (i.e., $a_i \to 0$), and also when $\xi_{\rm obs}$ is
large. 

We can continue by integrating (\ref{dVdtau}) over $\tau$, from
zero up to an arbitrary value, thereby determining the total
4-volume available for the nucleation of bubbles that could
collide with an observer located at $(\xi_{\rm obs}, \tau)$.  One
finds
\begin{eqnarray}
  V_4&=& {4 \pi \over 3} \left\{\tanh^2 \left( {\tau
     \over 2} \right) - \ln \left[ \epsilon \cosh^2 \left( {\tau
     \over 2} \right) \right] \right\} \nonumber \\ 
     && \quad - 4 \pi \left\{a_i (1-\epsilon) - {1 \over 2} a_i^2
     (1-\epsilon^2) + {1 \over 9} a_i^3 (1-\epsilon^3) \right\},
\label{V4int}
\end{eqnarray}
where
\beq
  \epsilon = {1 \over \cosh \xi_{\rm obs} \sinh \tau + \cosh \tau }.
\eeq
This quantity represents the 4-volume of the region that is in
the past light-cone of the point $(\xi_{\rm obs},\tau)$, but is
not inside the reference bubble nor in the past light-cone of its
nucleation event.  This volume can be calculated directly without
making the coordinate transformation (\ref{coordtran}), and we
have verified that the results agree.

If an observer at $\xi=\xi_{\rm obs}$ has not seen a bubble at
time $\tau_1$, then the probability that she will not be hit by a
bubble by some later time $\tau_2$ is determined by
(\ref{V4int}), with
\beq
P = \exp \left\{ - \lambda \left[ V_4(\xi_{\rm obs}, \tau_2) -
     V_4(\xi_{\rm obs}, \tau_1) \right] \right\} .
\eeq

It is particularly interesting to look at (\ref{V4int}) for large
$\xi_{\rm obs}$, since a typical observer will be found at
arbitrarily large $\xi_{\rm obs}$.  In that limit one finds that
\beq
  \epsilon = {2 e^{-\xi_{\rm obs}} \over \sinh \tau} - {4 \cosh
     \tau \over \sinh^2 \tau} e^{-2 \xi_{\rm obs}} + 
     O(e^{-3 \xi_{\rm obs}}) ,
\eeq
which is valid for any $\tau \not = 0$.  Then
\begin{eqnarray}
  V_4 &=& {4 \pi \over 3} \left\{{\xi_{\rm obs}+\tanh^2 \left({\tau
     \over 2}\right) + \ln \left[{\tanh \left({\tau \over
     2}\right)}\right]+2 e^{-\xi_{\rm obs}} \coth(\tau )}\right\}
     \nonumber \\
     && \quad - 4 \pi \left\{{a_i- {1 \over 2} a_i^2+ {1 \over 9}
     a_i^3- 2 a_i {e^{-\xi_{\rm obs}} \over \sinh \tau}}\right\}
     + O(e^{- 2 \xi_{\rm obs}}).
\end{eqnarray}
Then, if we are interested in the limit $\tau \rightarrow
\infty$, we find
\beq
  \lim_{\tau \rightarrow \infty} V_4 = {4 \pi \over 3} \left(
     \xi_{\rm obs} + 1 + 2 e^{-\xi_{\rm obs}} \right) - 4 \pi
     \left( a_i - {1 \over 2} a_i^2 + {1 \over 9} a_i^3 \right) +
     O (e^{- 2 \xi_{\rm obs}}) .
\eeq
The probability of a point not being hit by a bubble is given by
$P = \exp(- \lambda V_4)$, so the leading term $4 \pi \xi_{\rm
obs}/3$ in the formula above reproduces Eq.~(\ref{omegau}), which
was used to determine the fractal dimension.

\section{A moving observer in de Sitter}
\label{moving}

Even for an observer in the false vacuum, the bubble nucleation
rate and the angular distribution of bubbles depend on the
observer's velocity relative to the ``preferred'' co-moving
congruence ${\cal C}$ which is determined by the surface of
initial conditions. Here, we give a self-contained account of
this effect.

Consider a de Sitter space with $H=1$,
\beq
ds^2=\eta^{-2}(d\eta^2-d{\bf x}^2),
\label{desitter}
\eeq
with $-\infty<\eta<0$. The conformal time $\eta$ is related to
the usual time vatiable as
\beq
\eta=-e^{-t}.
\label{eta}
\eeq
We shall assume that there are no bubbles at some initial moment
$\eta=\eta_i$.

Consider an observer at ${\bf x}=0,~ \eta=\eta_0$, moving with a
velocity
\beq
v\equiv\tanh\phi
\eeq
relative to the comoving observers of (\ref{desitter}). We want
to know the probability for this observer to be hit by a bubble
per unit proper time (by his clock).

An infinitesimal proper time interval $\delta\tau$ corresponds to
a coordinate displacement
\beq
\delta\eta=|\eta_0|\delta\tau\cosh\phi,
\label{deltaeta}
\eeq
\beq
\delta {\bf x}={\bf v}\delta\eta.
\label{deltax}
\eeq
The probability to be hit by a bubble is determined by the
spacetime volume between the past light cones of the points
$(\eta_0,~0)$ and $(\eta_0+\delta\eta,~\delta {\bf x})$. The
first of these light cones is given by
\beq
|{\bf x}_1(\eta)|=\eta_0 -\eta
\label{cone1}
\eeq
and the second is given by
\beq
|{\bf x}_2 -{\bf \delta x}|=\delta\eta+\eta_0-\eta.
\label{cone2}
\eeq
To linear order in $\delta\tau$,
\beq
|{\bf x}-{\bf \delta x}|\approx r-|\eta_0|\delta\tau \sinh\phi
\cos\theta = r-v\delta\eta\cos\theta,
\label{approx}
\eeq
where $r=|{\bf x}|$ and $\theta$ is the angle between ${\bf x}$
and ${\bf v}$, so we can rewrite (\ref{cone2}) as
\beq
r_2(\eta,\theta)=\delta\eta(1+v\cos\theta)+\eta_0-\eta.
\label{cone2'}
\eeq

The spacetime volume between the two light cones is given by the
integral
\beq 
\delta V_4 =2\pi\int_{\eta_i}^{\eta_0}d\eta\eta^{-4}
\int_0^\pi d\theta \sin\theta r_1^2(\eta)\delta r(\theta)
=2\pi\delta\eta \int_{\eta_i}^{\eta_0} d\eta\eta^{-4}
(\eta_0-\eta)^2
\int_0^\pi d\theta\sin\theta (1+v\cos\theta),
\label{int}
\eeq
where $\delta
r(\theta)=r_2(\eta,\theta)-r_1(\eta)=\delta\eta(1+v\cos\theta)$.
The integration over $\eta$ is easily done by a change of
variable $\xi=-1/\eta$,
\beq
\int_{\eta_i}^{\eta_0} d\eta\eta^{-4} (\eta_0-\eta)^2 =\int d\xi 
(\eta_0\xi +1)^2 ={1\over{3}}\eta_0^2(\xi_0-\xi_i)^3.
\label{inteta}
\eeq
Substituting this into (\ref{int}) and using Eq.~(\ref{deltaeta})
for $\delta\eta$, we have
\beq
\delta V_4={2\pi\over{3}}\delta\tau\cosh\phi f(t_0-t_i)
\int_0^\pi d\theta\sin\theta (1+v\cos\theta),
\label{V4}
\eeq
where
\beq
f(t)=\left(1-e^{-t}\right)^3
\label{f}
\eeq
and $t$ is the usual time variable which is related to $\eta$ as
in Eq.~(\ref{eta}). 

Note that $f(t)\to 1$ as $t\to\infty$. In what follows we assume
the limit $t_i\to-\infty$ and set $f(t_0-t_i)=1$.

The $\theta$-integration in (\ref{V4}) is, of course, easily
done; the result is
\beq
\delta V_4={4\pi\over{3}}\delta t,
\label{V4total}
\eeq
where
\beq
\delta t=|\eta_0|^{-1}\delta\eta =\delta\tau\cosh\phi
\label{deltat}
\eeq
is the interval of time $t$ corresponding to the proper time
interval $\delta\tau$. 

The result (\ref{V4total}) is easy to understand. We have two
past light cones, one has its origin at $(t_0,0)$ and the other
has its origin shifted in both time and space directions.
Eq.~(\ref{V4total}) tells us that the 4-volume difference $\delta
V_4$ depends only on the time displacement $\delta t$. And this
is as it should be. The volumes of the light cones do not change
when we shift them in the ``horizontal'' (space) directions, so
$\delta V_4$ remains unchanged, as long as one light cone remains
entirely within the other.

All observers see the same bubble nucleation rate per unit time
$t$, but this corresponds to different rates per unit proper time
$\tau$,
\beq
{\delta V_4\over{\delta\tau}}={4\pi\over{3}}\cosh\phi.
\eeq
There is thus a preferred frame, where $\phi=0$, corresponding to
the lowest nucleation rate.

The distribution of the arrival directions of the bubbles is also
anisotropic. The angular distribution can be easily read from
Eq.(\ref{V4}),
\beq
{d(\delta V_4)\over{\delta\tau d\Omega}}={1\over{3}}\cosh\phi
(1+v\cos\theta). \label{result}
\eeq 

The velocity of a geodesic observer relative to the comoving
frame decays as
\beq
v(t)\propto e^{-t}
\eeq
at large $t$, so the anisotropy rapidly disappears and the rate
approaches that for a comoving observer. However, when we study
bubble collisions and construct FRW coordinates inside a bubble,
the set of comoving observers associated with these coordinates
includes observers with arbitrarily large values of $v$. This is
the origin of the anisotropy measured by typical observers in a
bubble universe.

The frame-dependence of the bubble arrival rate is perhaps not
too surprising.  To the moving observer the initial surface
$t=t_i\to -\infty$ looks rather odd.  It is a null surface which
crossed the worldline of the observer a finite proper time $\tau$
in the past. This time has been calculated in \cite{BGV}; it is
given by
$$
\tau = {1\over 2} \ln\left({\gamma + 1\over \gamma - 1}\right).
$$ 
For $(\gamma - 1) \ll 1$, the initial surface is many Hubble
times ago, and the difference from a comoving observer is small.
However, for $\gamma \gg 1$, the cutoff surface is only a small
fraction of the Hubble time away. No wonder it changes the rate
and introduces an asymmetry.

The connection between (\ref{result}) and the results of the
previous Section is not completely straightforward, and deserves
some comment. The main difference is that observers inside the
bubble will not have any bubbles nucleating in their immediate
neighborhood (since we assume that bubbles can only nucleate in
false vacuum). 

Near the surface $\tau=0$, corresponding to the origin of time in
the open chart, this difference becomes irrelevant.  For
$\tau=0$, Eq.~(\ref{lagr}) gives
\begin{equation}
{dV_4 \over d\tau d\Omega'} = {1\over 3 \gamma^3 (1-\beta\cos\theta')^3}.
\end{equation}
At first sight, this looks rather different from (\ref{result}).
However, if we take into account the aberration of the angles in
a moving frame,
$$
\cos\theta={\cos\theta'-\beta \over 1-\beta\cos\theta'},
$$ 
we find
$$ 
{dV_4 \over d\tau d\Omega} = {1\over
3}\gamma(1+\beta\cos\theta).  
$$ 
This agrees with (\ref{result}) under the identifications
$\gamma=\cosh\phi$ and $\beta=v$.  From Eq.~(\ref{deco}), the FRW
observer at constant $\xi$ has a relativistic factor with respect
to the preferred ``rest frame'' congruence ${\cal C}$ which is
given by
\begin{equation}
\gamma= {dt\over d\tau} ={\sinh\tau + \cosh\xi\cosh\tau \over \cosh\tau+\cosh\xi\sinh\tau}.
\end{equation}
For $\tau = 0$, we have $\gamma=\cosh\xi$, which is unbounded for
large $\xi$. Hence, right after nucleation, observers in this FRW
congruence have arbitrarily large relativistic factors.

For any $\tau>0$, the limit of very large $\xi$ gives a finite
$\gamma\approx \coth\tau$. This means that a period of inflation
inside the bubble slows down {\em all} observers to
non-relativistic speed (for $\tau \gg 1$, the velocity is of
order $e^{-\tau}$).  Nevertheless, to the observers at large
$\xi$, the initial surface looks ``slanted,'' sloping down in
time in the direction away from the origin. This leads to the
anisotropy in the distribution of bubbles.  In embedding
coordinates, the initial surface at $t= t_i$, is described as the
intersection of the null surface $W+V=e^{t_i}$ with the de Sitter
hyperboloid (\ref{theh}) [the coordinate $V$ is ``time,'' whereas
the $W$ axis is normal to the hyperboloid at the nucleation point
$(V=0,W=1)$]. Now, we can always bring an observer from very
large $\xi$ to the origin of coordinates $\xi'=0$ by means of a
large boost, with $\beta=\tanh\xi\approx 1$. If in the original
``rest'' frame the observer was at large $z$, the surface of
initial conditions in the new reference frame takes the form
$V'+Z'\approx 0$. This ``initial'' plane is (almost) tangent to
the bubble cone, along its null generator in the negative $z'$
direction. Since this leaves no room between the initial surface
and the reference bubble, it is clear that the probability of
being hit from that particular direction vanishes. From the point
of view of the rest frame ${\cal C}$ the anisotropy is not
surprising either.  The observers at large $\xi$ are very close
to the light-cone, near the surface of the reference bubble. 
Other bubbles can only hit from the false vacuum outside, and
because of that FRW observers are more likely to be hit by
bubbles approaching from even larger $\xi$ than from any other
direction.

\section{Conclusions}

Our primary goal in this project was to study the effect of
bubble collisions on the structure of ``bubble universes'' in the
inflating false vacuum. In particular, we wanted to know how
likely it is for an observer living in one of the bubbles to be
affected by such a collision.

In the absence of collisions, the bubble interior is described by
an open FRW model. A constant-FRW-time slice of such an
unperturbed bubble universe is a hyperboloid, a space of constant
negative curvature. Each bubble collision carves an infinite
wedge-like region out of the hyperboloid.  We found that the part
of the hyperboloid that remains unaffected by collisions has a
fractal character. Its volume is infinite, but it constitutes a
vanishing fraction of the total volume of the hyperboloid. 

If we pick a random observer in the unaffected region, then we
find that the typical distance $d$ from this observer to the
nearest bubble collision is given by
\beq
d \sim R\ln(1/\lambda).
\label{dR}
\eeq
Here $R$ is the curvature radius of the hyperboloid and $\lambda
\ll 1$ is the bubble nucleation rate (per Hubble volume per
Hubble time in the false vacuum). The origin of the logarithmic
dependence on $\lambda$ is that the volume grows exponentially
with distance on a hyperboloid. Current observational bounds on
the spatial curvature imply $R\gtrsim 10 H_0^{-1}$, where
$H_0^{-1}$ is the present Hubble radius, and Eq.~(\ref{dR})
yields $d > 10 H_0^{-1}$.

Assuming that we live in a bubble universe, we have also
estimated the probability for a collision to occur within our
observable range. This is given by
\beq
P_{coll}\sim 4\pi\lambda/H_0 R \lesssim \lambda.
\eeq
The bubble nucleation rate $\lambda$ is usually exponentially
suppressed, and thus the chance for us to observe a bubble
collision is rather remote. 

In the process of this investigation, we have uncovered a
remarkable fact, that the probability for an observer to be hit
by a bubble has a strong dependence on the arrival direction of
the bubble.  The origin of this effect can be traced to a simpler
setting, which does not involve bubble collisions. Consider a
geodesic observer who lives in false vacuum. We want to know the
probability for this observer to be hit by a bubble, per unit
time by his clock. We found that this probability depends on the
observer's velocity ${\bf v}$ relative to a certain preferred
``co-moving'' frame, which is set by the initial conditions at
the beginning of inflation. The bubble nucleation rate is minimal
for co-moving observers with ${\bf v}=0$, and the most probable
arrival direction for a bubble is that opposite to ${\bf v}$.

For a single observer, these effects will be present only for a
brief period of time. If the initial state of false vacuum is
specified on some spacelike hypersurface, the co-moving frame is
defined by the congruence ${\cal C}$ of geodesics orthogonal to
that surface. The velocity of a geodesic observer relative to
${\cal C}$ redshifts exponentially with time, so the bubble
nucleation rate rapidly becomes isotropic and approaches its
co-moving value. This is in accord with the widespread belief
that the initial conditions at the onset of inflation have no
lasting effect.

We found, however, that this folk wisdom does not apply when the
situation is described in terms of a FRW open universe inside the
bubble. The geodesic congruence corresponding to such a universe
includes geodesics with velocities arbitrarily close to the speed
of light relative to ${\cal C}$. Fast moving observers in this
congruence would initially detect an extremely high rate of
bubble hits. The short period of inflation inside the bubble
slows down the congruence to non-relativistic speeds, and the
rate approaches a constant on the hyperboloid (at large distances
from the origin). Nevertheless, the angular asymmetry in the
arrival directions of the bubbles remains.  The reason is that,
to an observer far from the origin, the initial surface looks
very anisotropic, sloping down in time in the radial direction
further away from the origin. In the unlikely event that we
detect a signature of a bubble collision in some direction in the
sky, we will be able to say that we are probably moving in that
general direction relative to the preferred congruence ${\cal
C}$.

Now that we know that the initial conditions at the beginning of
inflation have a lasting effect on the distribution of bubble
hits, one cannot help wondering what other effects they may have.
We hope to return to this issue in the future.

\section*{Acknowledgements}

The work of JG was supported in part by CICYT and DURSI. The work
of AV was supported in part by grant PHY-0353314 from The
National Science Foundation and by grant RFP1-06-028 from The
Foundational Questions Institute.  The work of AHG was supported
in part by the U.S. Department of Energy (D.O.E.) under
cooperative research agreement \#DF-FC02-94ER40818, and also by
the Kavli Foundation.

\section*{Appendix 1}

In this appendix, we repeat the calculation of the distribution
of collisions which we presented in Section IIa, but now instead
of looking around the origin we consider the vicinity of a point
$\xi_{\rm obs}\gg 1$.

It is useful to go to the boosted frame where the point of
observation is at the origin, as we did in Section II. 
Furthermore, we may take the limit $\gamma=\cosh\xi_{\rm obs} \to
\infty,\beta=\tanh\xi_{\rm obs}->1$ in the expression for the
initial surface (\ref{insur}),
$$ 
a'_i = -2{\cos\theta'+F' \over 1+2F'\cos\theta'+F'^2}.  
$$ 
We may take Eq.~(\ref{stpt}) as the starting point. Performing
the integral over $a'$ we have
\begin{equation}
{dV_4} = {1\over 3F'}\left[\left({1+F'\over
1-F'}\right)^3-\left({1-F'^2\over
1+2F'\cos\theta'+F'^2}\right)^3\right] dF'd\Omega'.
\end{equation}
Integrating over solid angle, we have
\begin{equation}
{dV_4} = {8\pi (3+2F'+3F'^2)\over 3(1-F')^3(1+F')}dF'.
\end{equation}
Finally, integrating over $F'$ we obtain
\begin{equation}
V_4(\xi',\tau,t_i\to-\infty,\xi_{\rm obs}\gg 1) = {4\pi\over
3}\left[{4F'\over 3(1-F')^2}-\ln{1-F'\over
1+F'}\right]_{-e^{-\tau}}^{F'(\xi',\tau)}.
\end{equation}
After some algebra, this greatly simplifies to
$$
V_4(\xi',\tau,\xi_{\rm obs}\gg 1)={4\pi\over
3}\left[2e^{\xi'}\sinh\xi'\tanh^2(\tau/2)+\xi'\right].  
$$ 
This can be compared with the distribution around the privileged
point $\xi_{\rm obs}=0$ given in (\ref{alan}).  The expressions
are rather similar. In particular the asymptotic expressions
($\tau\gg1$)
\begin{eqnarray}
V_4 &\approx& 4\pi\xi' + O(\xi'^2) \quad(\xi'\ll 1),\\ 
V_4 &\approx& {4\pi\over 3} e^{2\xi'} + O(\xi')\quad(\xi'\gg 1).
\end{eqnarray}
are the same.

Suppose we match the inflating phase to a standard cosmological
phase with scale factor given by $a(\tau)$. As the universe
decelerates, the horizon becomes larger and more bubbles come
into sight. We can then ask what is the rate at which bubbles
become visible per unit time to the observer at $\xi'=0$. This is
proportional to the nucleation rate $\lambda$, times the rate at
which the relevant 4-volume enters the backward light cone from
the point of observation:
\begin{equation}
{dN(\tau) \over d\tau} = \lambda\left. {\partial V_4 \over \partial\xi'}\right|_{\tau=\tau_e,\xi'=\xi'(\tau)} 
{d\xi'(\tau)\over d\tau}.
\end{equation}
Here $\tau_e$ is the time at the end of inflation, and the last
factor gives the rate at which the observable distance changes as
we look back to the surface where inflation ends:
$$
\xi(\tau)=\int_{\tau_e}^{\tau} {d\tau\over a(\tau)}.
$$
For instance, if inside the bubble the vacuum energy is zero, and
we go immediately into the curvature dominated regime
$a(\tau)=\tau$, then the rate at which we would see new bubbles
entering our horizon per unit proper time is given by
\begin{equation}
{dN(\tau) \over d\tau} = {4\pi\lambda\over 3}\left[{\tau\over 2}+{1\over \tau}\right].
\end{equation}
Since this grows without bound, we would be guaranteed to see
some bubbles sooner or later if we lived in this Minkowski
vacuum.

\section*{Appendix 2}

Here, we explore the possibility of setting up initial conditions
which do not break the residual $O(3,1)$ symmetry of the bubble. 
This is a formal exercise whose practical utility is unclear.
Bubbles nucleate at random points, and it is not possible to set
up initial conditions which preserve $O(3,1)$ symmetry for all of
the bubbles in the ensemble. Nevertheless, the attempt may be
illustrative.

\begin{figure}[ht]
\begin{center}\leavevmode  %
\epsfxsize=16.5cm\epsfbox{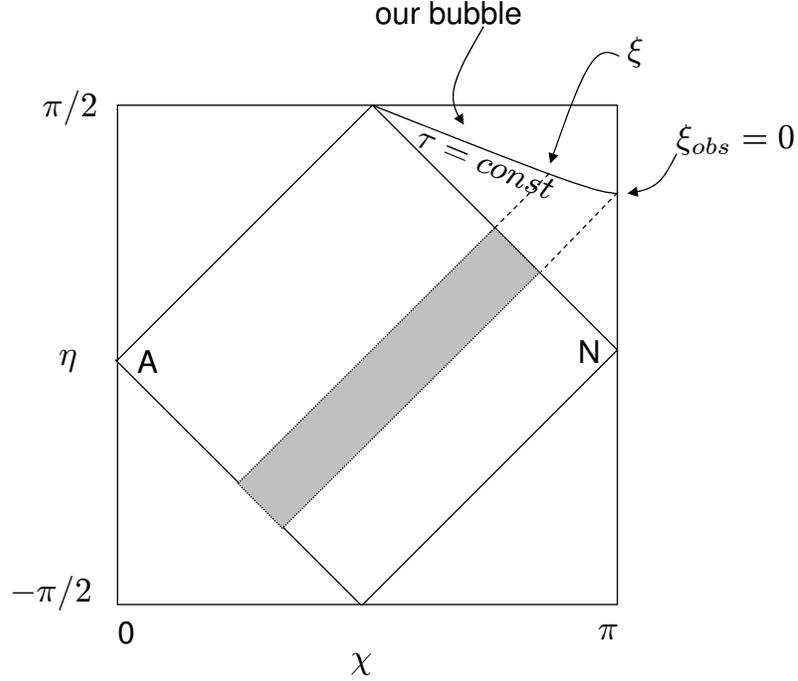}
\end{center}
\caption{Conformal diagram of a bubble in a de Sitter space. The relevant 4-volume to the past 
of a sphere of radius $\xi$ around the origin is shaded in grey.}
\label{2j}
\end{figure}

The simplest thing we can do in order to have $O(3,1)$ invariant
initial conditions, is to require that there are no bubbles
inside the backward light-cone from the antipodal point A of the
nucleation event N (see Fig. \ref{2j}).  We are also assuming
that N is in false vacuum, so the surface of initial conditions
is the disjoint union of the backward light cones from A and N. 
The relevant four-volume of our interest is represented by the
shaded area in Fig. \ref{2j}.

It is convenient to use the conformal closed chart, in which the
metric reads
\begin{equation}
ds^2= {1\over \cos^2 \eta}(-d\eta^2 + d\chi^2 + \sin^2\chi d\Omega^2).
\end{equation}
The 4-volume is given by
\begin{equation}
\tilde V_4(\xi,\tau) = 4\pi \int {\sin^2\chi \over \cos^4\eta} d\chi d\eta = 8\pi \int_{u_{min}}^{u_{max}} du \int_0^{\pi/2} dv
{\sin^2(v-u) \over \cos^4(v+u)},
\end{equation}
where we have introduced the change of coordinates $\eta= v+u$
and $\chi=v-u$. Past light cones are labeled by $u=const$ values.
Performing the integrals, we have
\begin{equation}
\tilde V_4(\xi,\tau)={8\pi\over 3}\left[{2\cos 2 u\over \sin^2 2 u} - \ln \tan^2 u\right]_{u_{min}}^{u_{max}}.\label{tb4}
\end{equation}
Finally, we must relate the values of $u$ to the values of $\xi$
on the $\tau=const.$ hyperboloid. Alternatively, we may relate
them to the by now familiar variable $F$.

On the bubble cone we have $\chi=\pi-\eta$, and hence the
physical radius of two-spheres is given by
\begin{equation}
R={\sin\chi\over\cos\eta}= \tan\eta=-\cot u.
\end{equation}
On the other hand, in terms of $a$ and $F$ coordinates we have
\begin{equation}
R=(1+aF) = {1+F\over 1-F},
\end{equation}
where we have used that on the bubble cone $a=2/(1-F)$ [see
Eq.~(\ref{tcrc})]. Equating both expressions for $R$ we have
\begin{equation}
\tan u = {F-1 \over F+1}.
\end{equation}
Substituting in (\ref{tb4}) we obtain
\begin{equation}
\tilde V_4 = {8\pi\over 3} \left[{4F(F^2+1)\over (F^2-1)^2} - 2 \ln {1-F\over 1+F}\right]_{-e^{-\tau}}^{F(\xi\tau)}.
\end{equation}
Note that the leading dependence for large $\xi$ is the same as
in $V_4$, given in (\ref{v444}). However, both expressions differ
in the subleading terms, and so the limit of large $\xi_{\rm
obs}$ studied in the previous Section does {\em not} exactly
agree with the result which we have just obtained by setting up
Lorentz invariant initial conditions.

Formally, we have obtained a finite result which does not break
the homogeneity and isotropy of the reference bubble, by assuming
that the point of observation is still in false vacuum. However,
it is clear that the 4-volume $\tilde V_4$ diverges when we set
$\tau=0$ in the lower limit of integration . What this means is
that the bubble has probability 1 of being hit by other bubbles
immediately after its formation. In other words, there is no
possibility of eternal inflation with these initial conditions:
we have included too much of the contracting part of de Sitter
space.

Finally, we could have adopted yet another approach which
formally does not require any ``cut-off'' initial surface.
Indeed, we could ask the following question. Given that a
particular point has not been hit by a bubble at some late time
in the true vacuum (e.g. by the time of last scattering), what is
the expected distance to the nearest bubble? If the true vacuum
is of sufficiently low energy, the relevant 4-volume is finite,
without the need of any cut-off. This is illustrated in Fig. 3.
For clarity, we draw the case where the ``true'' vacuum inside
the bubble is Minkowski. In practice, it is enough that the true
vacuum be of sufficiently low energy density, so that the
observer's cosmological horizon is much bigger than the Hubble
size during inflation. In this case, the shaded area
corresponding to the 4-volume available for the nucleation of
bubbles which will hit a distance $\Delta\xi$ away from the point
of observation, is finite. Formally, we avoid the need to specify
the initial surface.

Again, this formal set-up is unphysical. {\em Some} initial
conditions must be specified, and these will look different from
the point of view of different observers. This will lead to the
anisotropies which we have discussed in the present paper.

\begin{figure}[ht]
\begin{center}\leavevmode  %
\epsfxsize=16.5cm\epsfbox{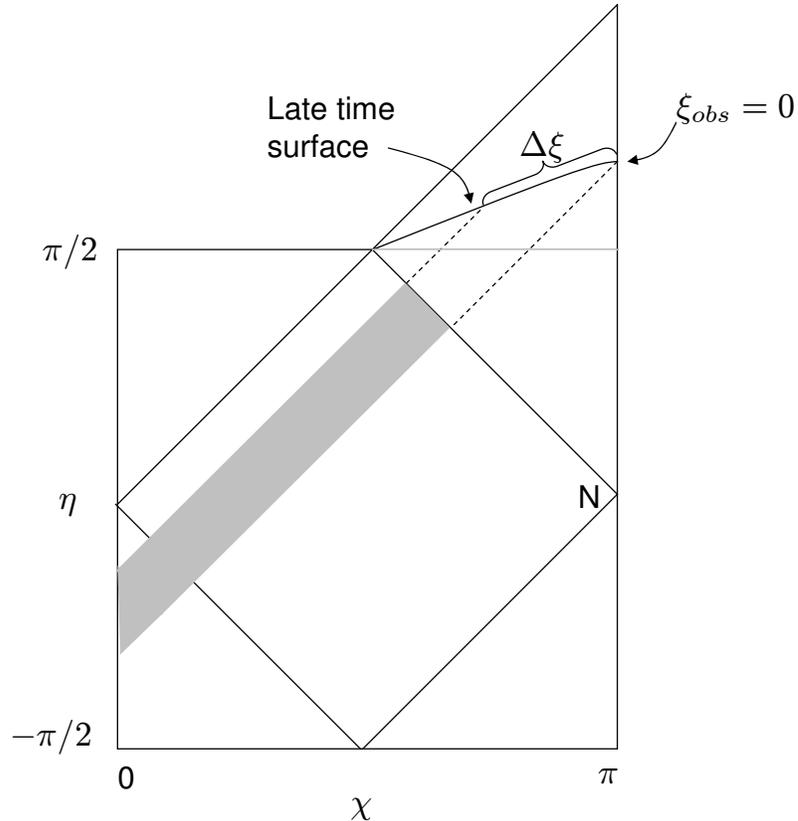}
\end{center}
\caption{Same as in Fig. \ref{2j}, but now the ``true'' vacuum inside the bubble 
is of lower energy. For simplicity, we take it to be Minkowski,
although this is not essential. In this case, the shaded area
corresponding to the 4-volume available for the nucleation of
bubbles which will hit a distance $\Delta\xi$ away from the point
of observation, is finite.  All we need is that the backward
light cone from the point of observation reaches the point
$\chi=0$ at some time in the past. This will happen if the
cosmological horizon at the time of observation is many times
larger than the Hubble size during inflation. Formally, we do not
need a cut-off initial surface, and seemingly we obtain a finite
answer which respects $O(3,1)$ invariance. However, we know that
physically some initial conditions are needed. These look
different to different observers in the FRW congruence, which
leads to anisotropies in the distribution of bubbles.}
\label{3j}
\end{figure}

\end{document}